\documentclass[runningheads,a4paper]{llncs}
\usepackage{amsmath}
\usepackage{amssymb}
\usepackage{graphicx}
\usepackage{cite}
\usepackage{todonotes}
\usepackage{subfig}
\usepackage{float}
\usepackage{algorithm}
\usepackage[noend]{algpseudocode}

\graphicspath{{pics/}}

\usepackage{url}
\urldef{\mailsa}\path|{isomurodov,loboda, alserg}@rain.ifmo.ru|
\newcommand{\keywords}[1]{\par\addvspace\baselineskip
\noindent\keywordname\enspace\ignorespaces#1}

\begin{document}
\mainmatter  % start of an individual contribution

\title{Ranking vertices for active module recovery problem}

\titlerunning{Ranking vertices for active module recovery problem}

\author{Javlon E. Isomurodov\inst{1,2} \and Alexander A. Loboda\inst{1,2}%
 \and Alexey A. Sergushichev\inst{1,2}}
\authorrunning{
Ranking vertices for active module recovery problem}


\institute{Computer Technologies Department, ITMO University, 
    Saint Petersburg, Russia \\
\and 
    JetBrains Research, Saint Petersburg, Russia\\
\mailsa
}

% search for the string "\mainmatter" where a contribution starts).

\toctitle{Lecture Notes in Computer Science}
\tocauthor{Authors' Instructions}
\maketitle

\begin{abstract}
    Selecting a connected subnetwork enriched in individually important 
    vertices is an approach commonly used in many areas of bioinformatics, including
    analysis of gene expression data, mutations, metabolomic profiles and 
    others. It can be formulated as a recovery of an active module from
    which an experimental signal is generated. Commonly, methods for solving
    this problem result in a single subnetwork that is considered to be a good
    candidate. However, it is usually useful to consider not one but multiple
    candidate modules at different significance threshold levels.
    Therefore, in this paper we suggest to consider a problem of finding 
    a vertex ranking instead of finding a single module.
    We also propose two algorithms for solving this problem: one that 
    we consider to be optimal but computationally expensive for real-world
    networks and one that works close to the optimal in practice and is also
    able to work with big networks.
\keywords{active module, vertex ranking, dynamic programming, 
integer linear programming, connected subgraphs}
\end{abstract}

\section{Introduction}
\label{sec_intro}

% :ToDo: refine

Network analysis has many applications in bioinformatics. This includes
analysis of co-expression network for gene clustering~\cite{Langfelder2008},
searching for reporter metabolites for metabolic processes~\cite{Patil2005},
or stratification of tumor samples based on topological distance 
between somatic mutations in a gene interaction networks~\cite{Hofree2013}.
The overall idea is that by taking into account interactions between entities
(genes, metabolites, etc.) one can better interpret the corresponding
raw data (gene expression, metabolite concentrations, etc.).

One type of network analysis corresponds to the active module recovery
problem. The goal of these methods is to find a connected 
subnetwork (module) that is enriched in individually important vertices. 
Such module, for example, could correspond to a signalling pathway for protein-protein
interaction network~\cite{Dittrich2008a} or a metabolic pathway for metabolic
networks~\cite{Jha2015}.

There are many implementations for active module 
recovery~\cite{Ideker2002,Dittrich2008a,Alcaraz2012,Sergushichev2016}.
These methods share a problem 
of non-monotonous dependence of the resulting module on the arbitrary
significance threshold value. This means that when a method is rerun 
with a more relaxed threshold not only some vertices can appear, but they can disappear
too. This situation is confusing for the user and makes
interpretation of the results harder.

In this paper we consider a formulation of the active module problem
in terms of connectivity-monotonous vertex ranking. This allows to 
generated modules for multiple thresholds that are consistent with each other.
First, 
in section~\ref{sec_formal_defs} we formally define the problem
and give related definitions. Then, in sections~\ref{sec_optimal}
and~\ref{sec_semiheuristic} we propose two methods to solve the problem: 
a brute-force-based method and semmi-heuristic method based
on solving a series of integer linear programming (ILP) problems. 
We also define two baseline methods in section~\ref{sec_baseline}.
Finally in section~\ref{sec_experiments} we compare the methods 
with each other and baseline methods on generated and real networks.

\section{Methods}
\label{sec_methods}

\subsection{Formal definitions}
\label{sec_formal_defs}

In this section we give a formal definition of the active module recovery
problem in its ranking variant. Here we consider only networks with a simple
structure of an undirected graph.  

Let $G = (V, E)$ be a connected undirected graph and $w: V \to [0, 1]$ 
to be a weight function defined on its vertices. 
There is also an unknown connected subgraph (\emph{active module}) and 
corresponding set of vertices $M$.
Weights $w$ are assumed to be random variables such that 
vertices from $M$ are i.i.d. and follow a "signal" distribution
and vertices from $V \setminus M$ are also i.i.d. but follow a "noise"
distribution. 
Here we consider weights to be corresponding to P-values of a statistical test,
where null hypothesis holds for vertices from $V \setminus M$ and 
corresponding weights follow uniform distribution $U(0, 1)$.
Following~\cite{Dittrich2008a} vertices from $M$ are assumed to follow
a beta-distribution $B(\alpha, 1)$ for some parameter~$\alpha$.

\begin{definition}
    Let $G = (V, E)$ be a graph. A \emph{vertex ranking} of $G$ is a permutation of its
    vertices $V$. For a ranking $r = (r_1, r_2, \ldots, r_{|V|})$ 
    we consider vertices at the beginning of $r$ (e.g. $r_1, r_2, \ldots$) to
    be more \emph{important} and ranked higher than vertices 
    at the end (e.g. $r_{|V|}, r_{|V|-1}, \ldots$).
\end{definition}

\begin{definition}
    Let us call a vertex ranking $r$ of a connected graph $G$ to be 
    \emph{connectivity-monotonous}, if all subgraphs $G_k$
    induced by vertices from ranking prefixes $r_{1..k} = (r_1, \ldots, r_k)$ 
    for $k \in {1..|V|}$ are connected.
\end{definition}

For convenience we will consider a rank prefix $r_{1..k}$ as
a set $\{r_1, \ldots, r_k\}$ rather than a vector if the context requires it.

In this paper we will use AUC (Area Under the Curve) measure 
to define which ranking $r$ of graph $G$ better recovers module $M$.

\begin{definition}
    AUC value of a vertex ranking $r$ for graph $G = (V,
    E)$ and module $M \subset V$ can be calculated using formula: 
    \[ AUC(r|M) = \sum_{i=1}^n \left( 1 - \frac{|r_{1..i} \setminus M|}{|V
    \setminus M|} \right) \frac{[r_i \in M]}{|M|}, \]
    where $[r_i \in M]$ is equal to 1 if $r_i \in M$ and 0 otherwise.
\end{definition}

To summarize we define the considered problem as follows.

\begin{definition}
    Given a connected graph $G$, an unknown active module $M$ and
    vertices weights $w$ that follow beta- and uniform distributions for vertices
    from $M$ and $V \setminus M$ correspondingly, the ranking variant of the 
    active module recovery problem consists in finding a
    connectivity-monotonous
ranking $r$ with the maximal value of $AUC(r|M)$.
\end{definition}

Later in this paper we consider the parameter $\alpha$ of 
the beta-distribution $B(\alpha, 1)$
to be known. Similarly to~\cite{Dittrich2008a} one can infer parameters of the
beta-uniform mixture from the vertex weights using maximum likelihood approach.

\subsection{Optimal-on-average ranking}
\label{sec_optimal}

In this section we describe a method that finds ranking with 
the maximal expected value of AUC. Correspondingly, we call it 
\emph{optimal-on-average} method.

First, let consider a set $D \subset 2^V$ of all vertex sets that
induce a connected subgraph of $G$
and a discrete probability $P(M)$ defined for all $M \in D$. Together
this constitutes a probability space $\mathcal{M}$.

Our task is to find a ranking $r$ with the maximal
expected value of AUC score given a vector of vertex weights $w$:
\begin{equation} \label{formula:eauc} 
    E[AUC(r|\mathcal{M})] = \sum_{M \in D} P(M|w) \cdot AUC(r|M).
\end{equation}

% :ToDo: make multiline
A conditional probability of a module $P(M|w)$ can be calculated
using the Bayes' theorem:
\begin{multline}
    P(M |w) = \frac{P(w|M) \cdot P(M)}{P(w)}  \\
    = 
    \frac{P(M)}{P(w)} \cdot 
    \prod_{v \in M} B(\alpha,1)(w(v)) \cdot \prod_{v \in {V \setminus M}} U(0,1)(w(v)).
\end{multline}

Let us rewrite the formula~\ref{formula:eauc}:
\begin{multline}
    E[AUC(r | \mathcal{M})] = \sum_{M \in D} p(M|w) \sum_{i=1}^n \left(
1 - \frac{|r_{1..i} \setminus M)|}{|V \setminus M|}\right) \frac{[r_i \in
M]}{|M|} \\ =  \sum_{i=1}^n \sum_{M \in D}  \left(
    1 - \frac{|r_{1..i}\setminus M|}{|V \setminus M|}\right) \cdot \frac{p(M|w)
    \cdot [r_i \in M]}{|M|}.
\end{multline}

This allows us to calculate $E[AUC(r| \mathcal{M})]$ iteratively:
\begin{multline} \label{eqn:edp}
    E[AUC(r_{1..k}|\mathcal{M})] = E(AUC(r_{1..k-1}|\mathcal{M})) + \\
    \sum_{M \in D | r_k \in M} \left( 1 - \frac{|r_{1..k} \setminus M|}{|V \setminus M|}\right) \cdot \frac{p(M|w)}{|M|}. 
\end{multline}

Formula \eqref{eqn:edp} allows to calculate every $r_{1..k}$ prefix ranking
only one time.

This can be used to find the best ranking as shown
in the algorithm~\ref{alg:dp}. There we fill in an
array that for every set of vertices $D[i]$ from $D$ contains
a pair of values $dp[i].auc$ -- expected AUC value of the best 
connectivity-monotonous ranking of vertices $D[i]$
and $dp[i].ranking$ -- the corresponding ranking.
The function $getArea()$ calculates the second summand
of formula~\eqref{eqn:edp}.

%TODO: rewrite dp's type
\begin{algorithm}
    \caption{Optimal-on-average ranking}\label{alg:dp}
    \begin{algorithmic}[1]
        \Procedure{OptimalRanking}{$V,E$}
        \State $D \gets getConnectedSubgraphs(V, E)$ 
            \Comment{elements of $D$ ordered by size}
            \State $dp[D]$ : (auc: Double, ranking: Vector)
        \For{$i=1$ to $|D|$}
            \State $M \gets D[i]$
            \ForAll{$v \in M$}
                \If{$isNotConnected(M \setminus \{v\})$}
                    \State continue
                \EndIf
                \State $j \gets$  get index of $M \setminus \{v\}$ in $D$
                \State $auc \gets dp[j].auc + getArea(D, dp[j].ranking, v)$
                \If{$auc > dp[i].auc$}
                \State $\bar{v} \gets (dp[j].ranking, v)$
                \State $dp[i] \gets (\bar{v}, auc)$
                \EndIf
            \EndFor
        \EndFor

        \Return$dp[|D|].ranking$
        \EndProcedure
    \end{algorithmic}
\end{algorithm}

The time complexity of the algorithm~\ref{alg:dp} is $O(n^2 \cdot |D|^2)$.
One call to $getArea()$  requires $O(n \cdot |D|)$ time and it is multiplied
by $O(n \cdot |D|)$ for the outer loops. 

\subsection{Semi-heuristic ranking}
\label{sec_semiheuristic}

In this section we describe another approach for the vertex ranking
problem. This approach is inspired by BioNet method~\cite{Dittrich2008a}
and consists in solving a series of integer linear programming (ILP) problems
using IBM ILOG CPLEX library.
Compared to the optimal-on-average approach from the previous section this
method allows finding a ranking for large graphs in a rather 
reasonable time. As this method does not explicitly optimizes AUC score
we call this method \emph{semi-heuristic}.

First, similar to BioNet, let us find a subgraph of $G$ that is most likely to be
the active module. The most likely subgraph has the best (log)-likelihood
score. The log-likelihood score of the module can be calculated
as a sum of log-likelihood scores of the individual vertices in the module,
where individual score for vertex $v$ is calculated as:
$$ score(v) = \log \mathcal{L} (\alpha, 1|w(v)) = \log(\alpha \cdot {w(v)}^{\alpha -1}).$$

Now, we can find a connected subgraph $M$ with a maximal sum of vertex scores.
This corresponds to an instance of Maximum-Weight 
Connected Subgraph problem (MWCS). This problem is NP-hard but it can be 
reduced to an ILP problem and solved by IBM ILOG CPLEX as, for example, 
in~\cite{El-Kebir2014}.

Using the found subgraph $M$ we can define a crude partial ranking by saying that
vertices of $M$ go before $V \setminus M$.

Next, we define a procedure to refine such partial ranking. This
procedure takes two sets of vertices: a set $R$ that contains already ranked
vertices and a set $C$ that contain set of candidate vertices to be ranked. Then
we find a subset $X$ of $C$, so that $R \cup X$ is a connected and vertices from
$X$ should be ranked higher than $C \setminus X$.

Using this procedure we can recursively refine ranking up to the 
individual vertex level. Initially we solve an instance where $R$
is set to an empty set and $C$ contains all vertices. Then 
we do ranking for $(R, X)$ and $(R \cup X, C \setminus X)$.
We stop recursion when the candidate set consists of only one vertex.

A parameter of this procedure is how to select set $X$. For this end,
similarly to the first step, we solve an MWCS instance, but 
with an additional constraint that requires the solution to 
contain at least one vertex from $R$ and at least one but not all vertices
from $C$. We set $X$ as an intersection of the solution and the set $C$.
The corresponding instance is solved by a modified solver from~\cite{Loboda2016},
where corresponding constraints were added into the ILP formulation.

Overall algorithm is shown as algorithm~\ref{alg:shmyak}. 
The procedure $findMaximumSG()$ solves MWCS with the described additional 
constraints and returns chosen subset of vertices from $C$.
If $list$ size is more than one, we call $refineRanking()$ to get
a ranking of this set.  The algorithm returns a ranking $r$ of vertices $C$.

\begin{algorithm}
    \caption{Semi-heuristic ranking refinement}\label{alg:shmyak}
    \begin{algorithmic}[1]
        \Procedure{refineRanking}{$V, E, R, C$}
        \State $r$ : Ranking
        \While{$C.size != 0$}
            \State $list \gets  findMaximumSG(V, E, R, C)$
            \If{$list.size > 1$}
                \State $list \gets refineRanking(V, E, R, list)$
            \EndIf
            \State $r.addAll(list)$
            \State $R.addAll(list)$
            \State $C.removeAll(list)$
        \EndWhile
        \Return $r$
        \EndProcedure
    \end{algorithmic}
\end{algorithm}

\subsection{Baseline methods}
\label{sec_baseline}

As base line for the experiments we consider the following two methods.

The first method ranks vertices by their weights: the smaller the weight,
the higher is rank. This ranking is 
not connectivity-monotonous but is a good starting point. We will
call this method non-monotonous.

The second method consists in running BioNet algorithm for ten different
significance thresholds. As the BioNet modules ($M_1$, $M_2$, \ldots, $M_{10}$)
can be non-monotonous we use the following combining procedure. 
We assign the highest rank to vertices from $M_1$, the second highest
to $M_2 \ M_1$, the third to $M_3 \setminus (M_1 \cup M_2)$ and so on.
The significance thresholds are selected to be distributed at equal steps
between maximum and minimum log-likelihood vertex scores.

\section{Experimental results}
\label{sec_experiments}

We carried three series of experiments for different graph sizes.
First, we considered small graphs of about 20 vertices where we were able to thoroughly
compare all the considered methods. Next, we analyzed medium-sized graphs
of 100 vertices. For such sizes that are closer to the real-world ones
we analyzed all methods except optimal-on-average one, as it became 
computationally infeasible to run. Finally, we tested methods
on a real-world graph of two thousand vertices.

\subsection{Small graphs}

In the first experiment we have generated 32 different graphs of size 18.
Then an active module of size 4 was chosen uniformly at random.
Value of $\alpha$ was chosen from $U(0, 0.5)$ distribution.
Vertex weights were generated from corresponding beta- and uniform 
distributions.

\begin{figure}
    \centering
    \begin{tabular}{@{}cccc@{}}
        \includegraphics[width=6cm]{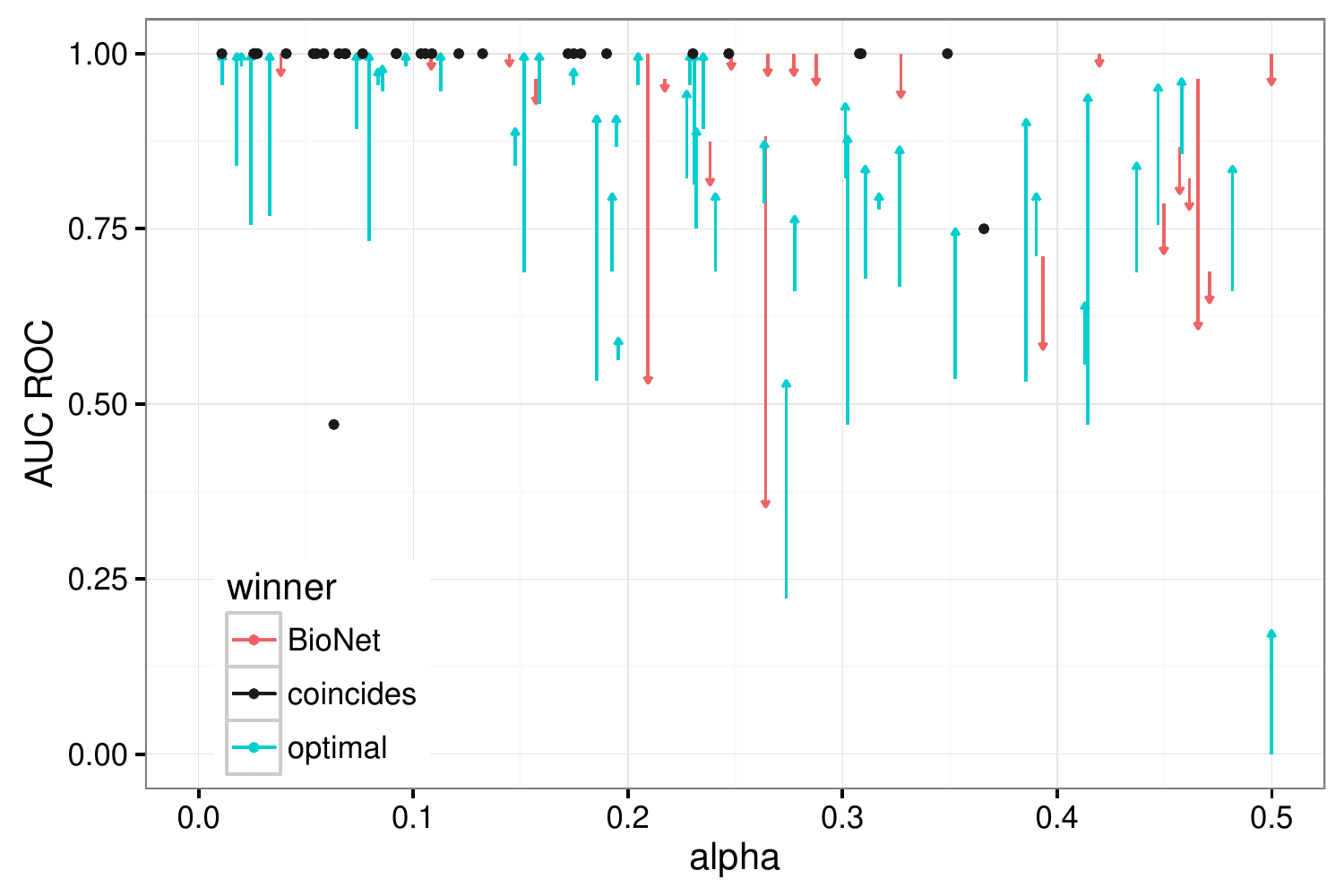} 
        &
        \includegraphics[width=6cm]{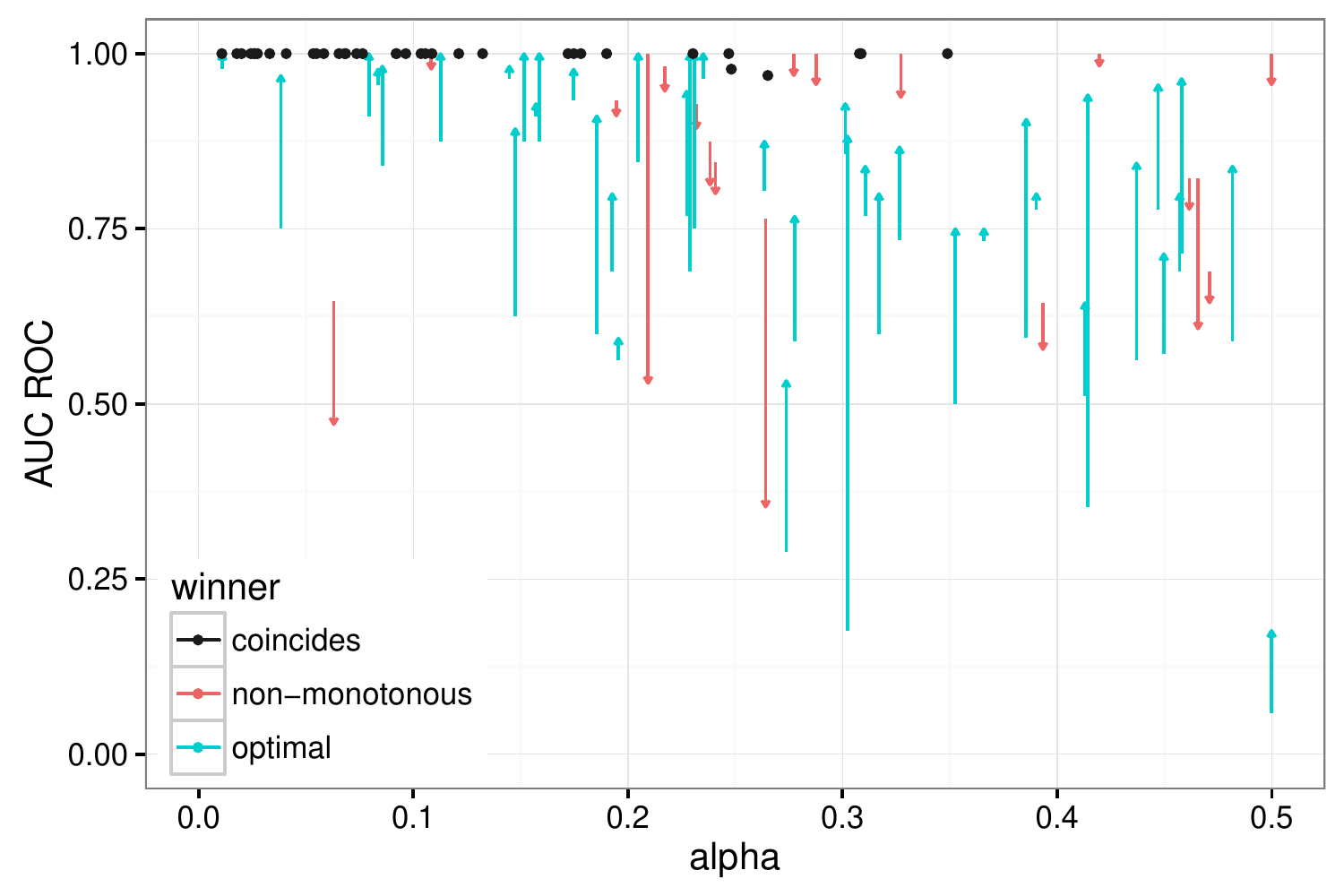} 
        \\
        \includegraphics[width=6cm]{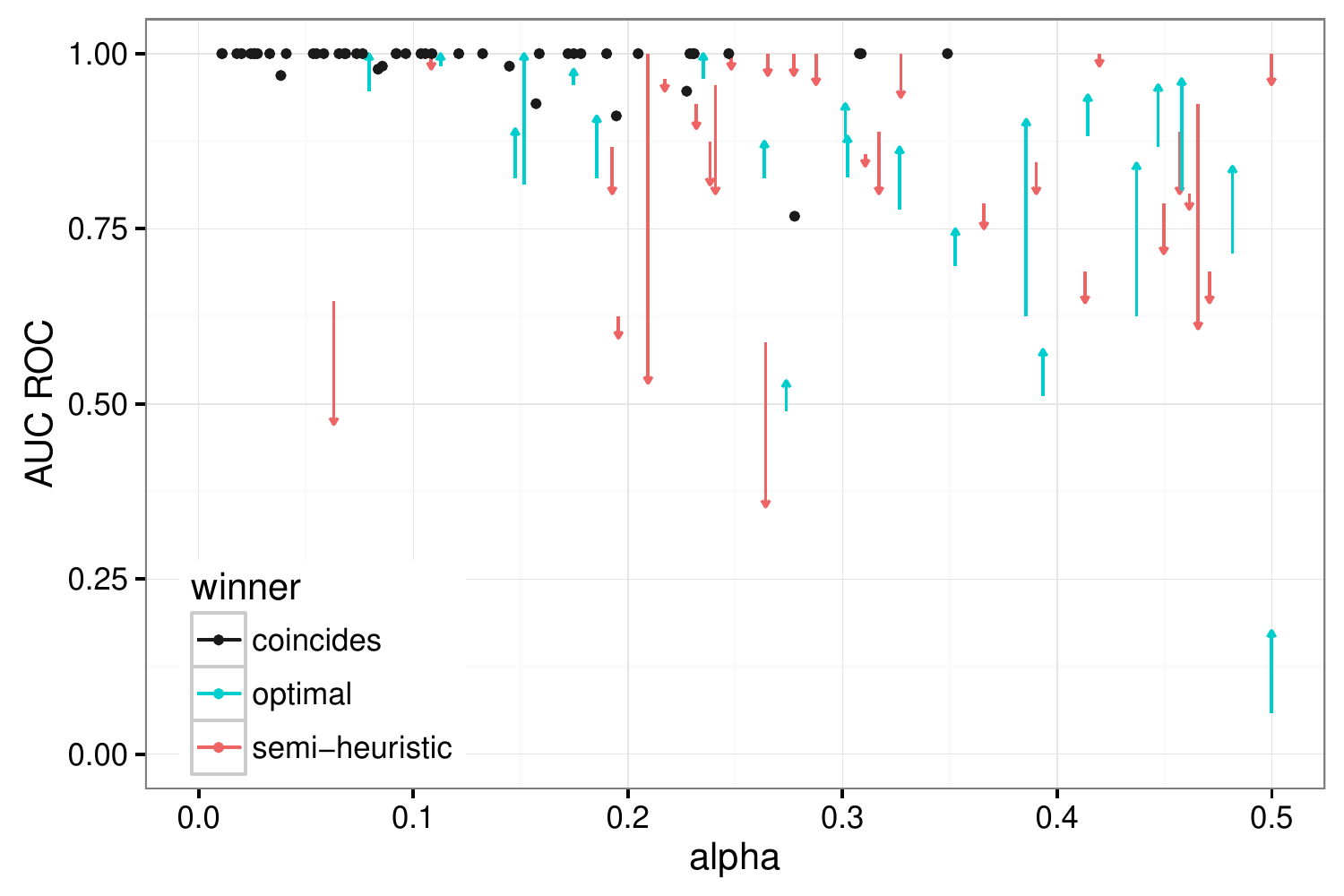} 
        &
        \includegraphics[width=6cm]{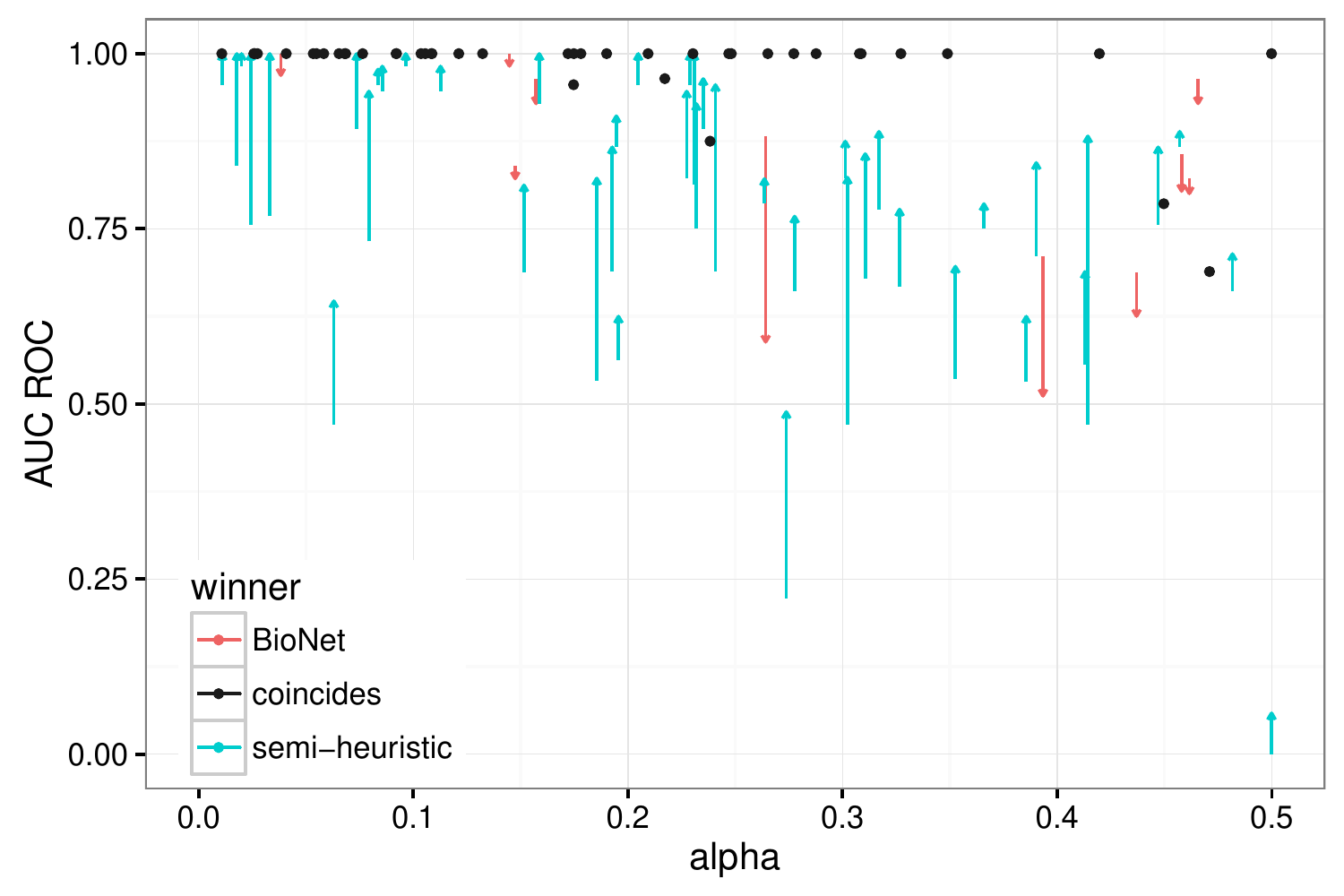}    
    \end{tabular}
    \caption{
        Module AUC values for graphs of size 18. The following methods are present:
    optimal-on-average, semi-heuristic, BioNet-like and non-monotonous.
    Each panel shows comparison of two methods. One arrow 
    correspond to one experiment with its ends corresponding to AUC value
    of the first and the second methods in the pair. The color depends
    on which method works better. True active modules were sampled from
    the uniform distribution.}%
    \label{fig:smalluni}%
\end{figure}

The results of the first experiment are shown on Fig.~\ref{fig:smalluni}.
They show that the optimal-on-average method in most cases works equal or better 
compared to both BioNet-like and non-monotonous baseline methods (top panels).
The semi-heuristic method works similarly well compared to optimal (bottom-left panel)
and better than BioNet-like method.

The distribution of active modules can be non-uniform in the real-world
data, so we also carried out an experiment with such non-uniform distribution
(see~\ref{sec_details} for details). Aside from the four methods
considered before we ran an optimal-on-average method parametrized by
the real empirical distribution of the modules. 

\begin{figure}
    \centering
    \begin{tabular}{@{}cccc@{}}
        \includegraphics[width=6cm]{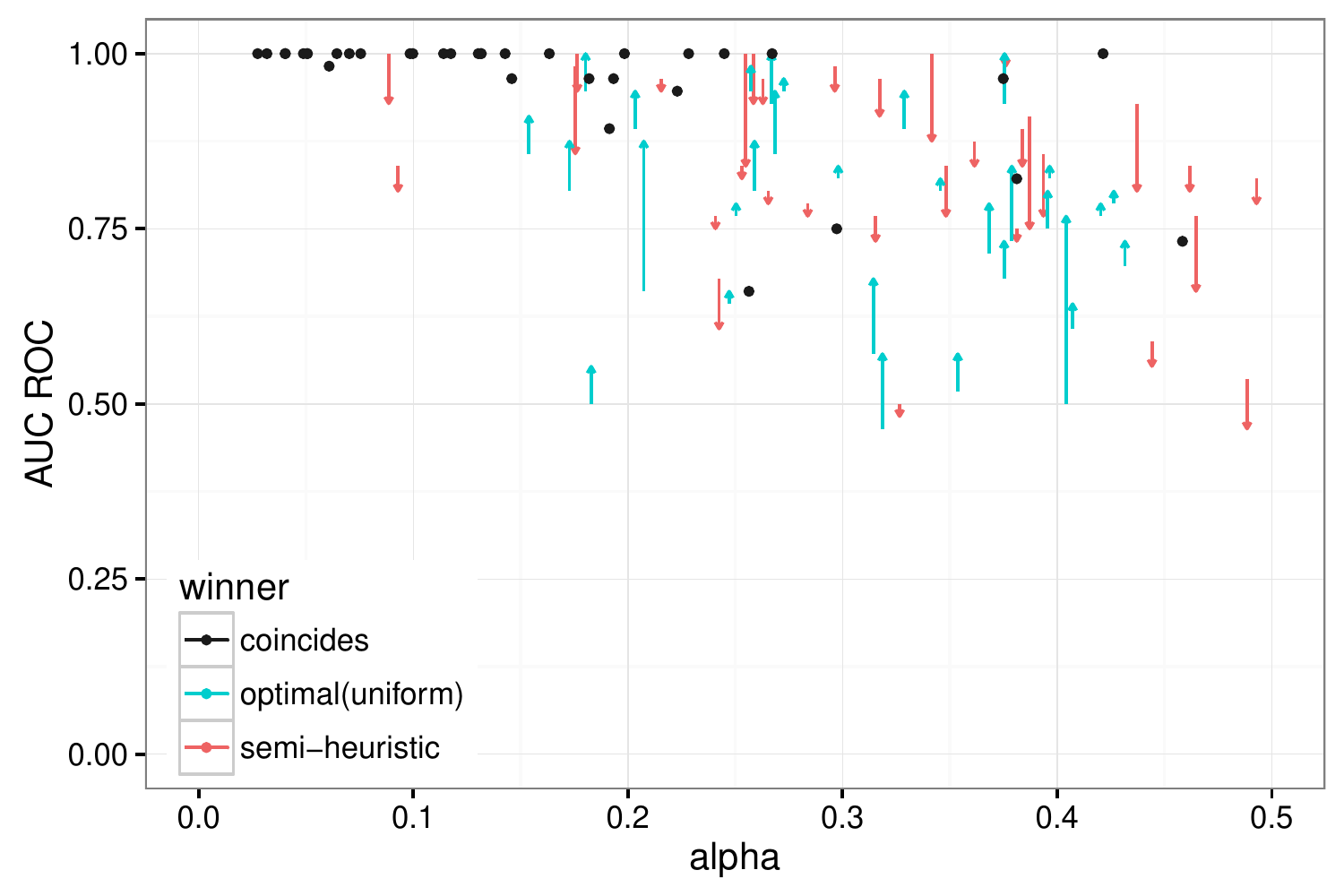} &
        \includegraphics[width=6cm]{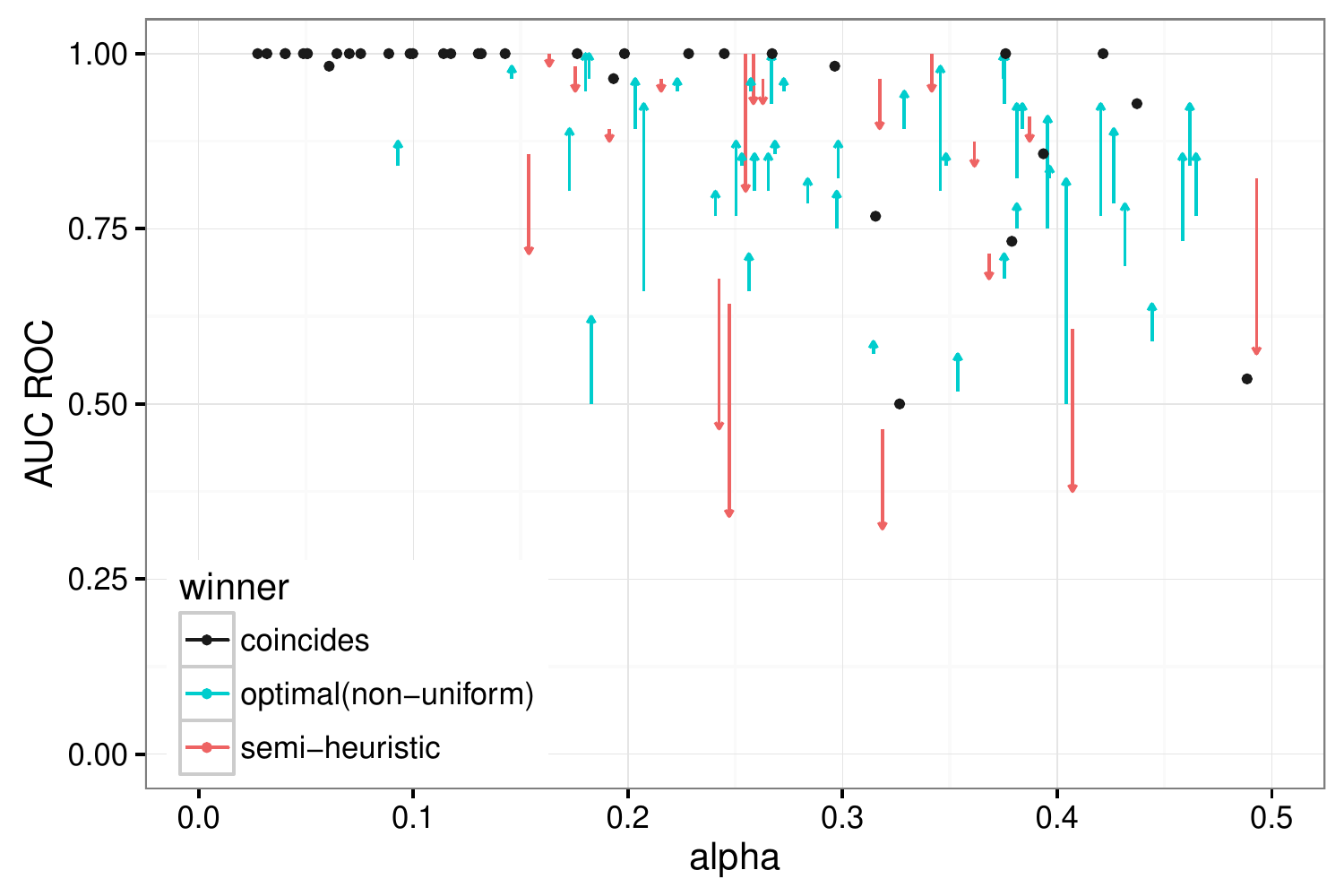}   \\ 
        \includegraphics[width=6cm]{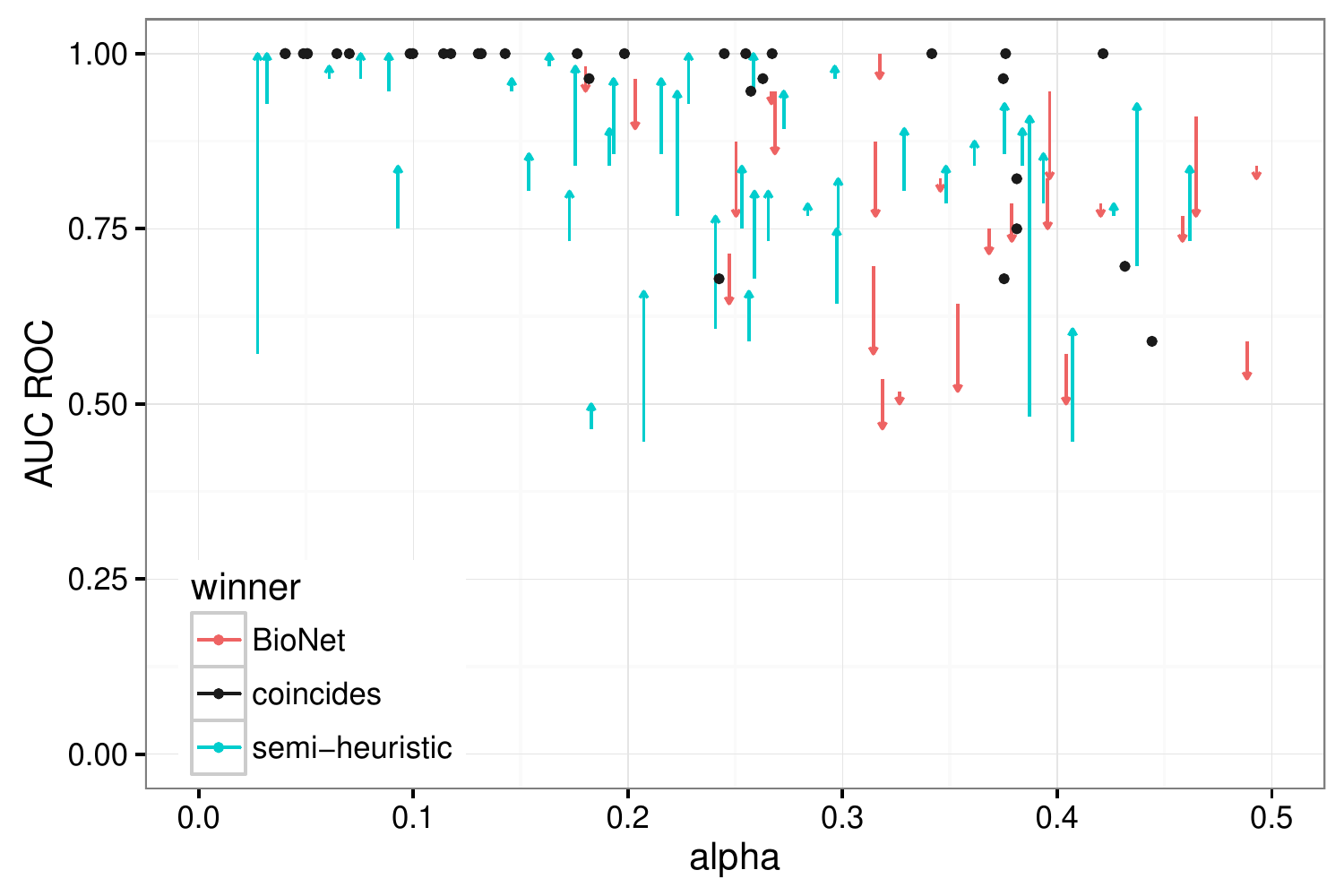} &
        \includegraphics[width=6cm]{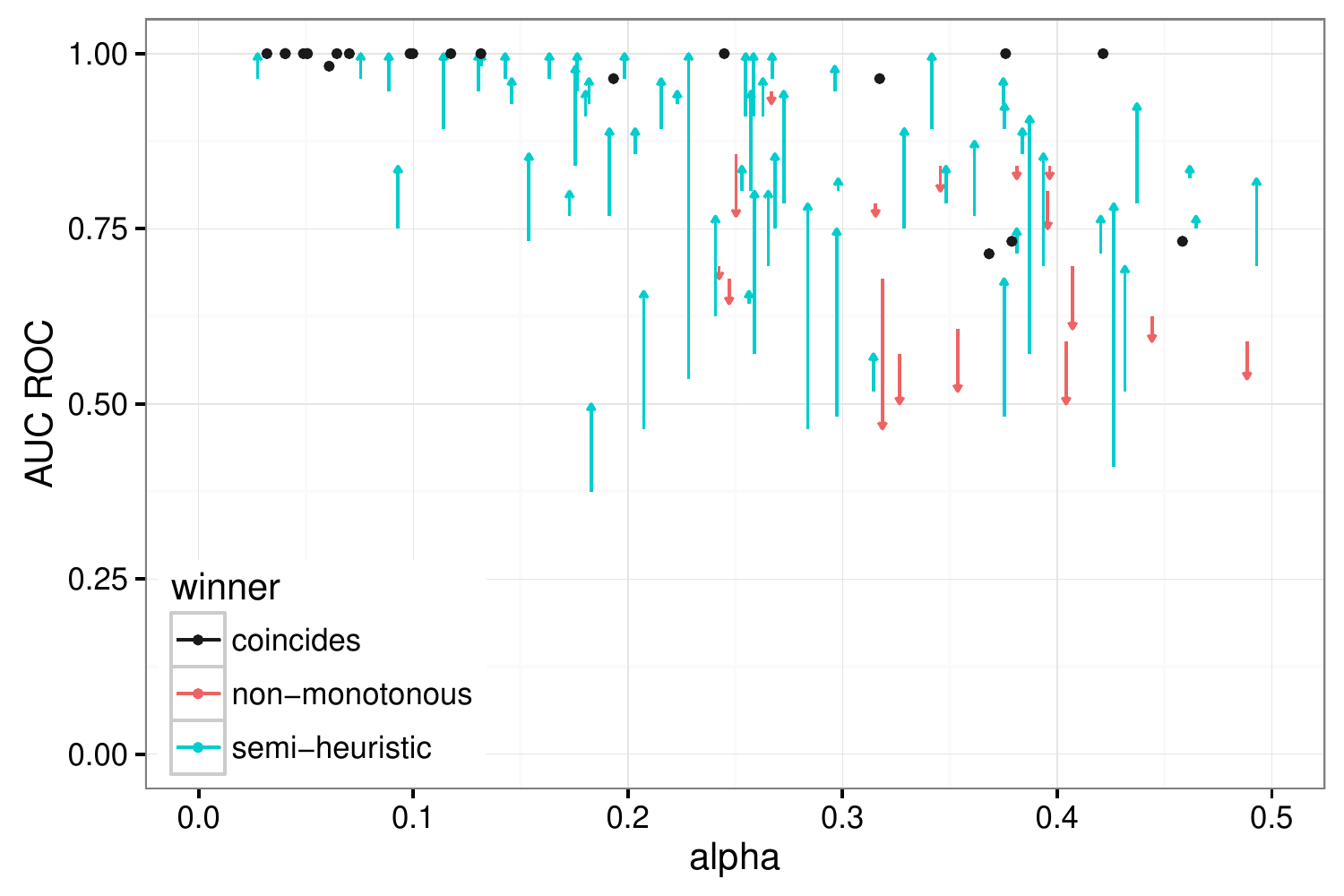}    
    \end{tabular}
    \caption{
        Module AUC values for graphs of size 18 when 
    true active modules were sampled from
    a non-uniform distribution.
        The following methods are present:
    optimal-on-average, optimal-on-average parametrized by the real distribution,
    semi-heuristic, BioNet-like and non-monotonous.
}%
    \label{fig:smallnon}%
\end{figure}

The results of this experiment are shown on Fig.~\ref{fig:smallnon}.
The situation is similar to the previous experiment with 
semi-heuristic method being close to optimal-on-average method
and better than baseline methods. However, the semi-heuristic method
works worse than optimal-on-average method parametrized by
the real modules distribution.

\subsection{Medium-sized graphs}

Similarly to the previous section we have generated 32 different graphs of size 100.
An active module were sampled to be the size of 5--25.

On these graph sizes running the optimal-on-average method becomes 
infeasible, so we excluded it from the analysis. 
A median time of running the semi-heuristic method 
was 146 seconds.

\begin{figure}
    \centering
    \begin{tabular}{@{}cccc@{}}
        \includegraphics[width=6cm]{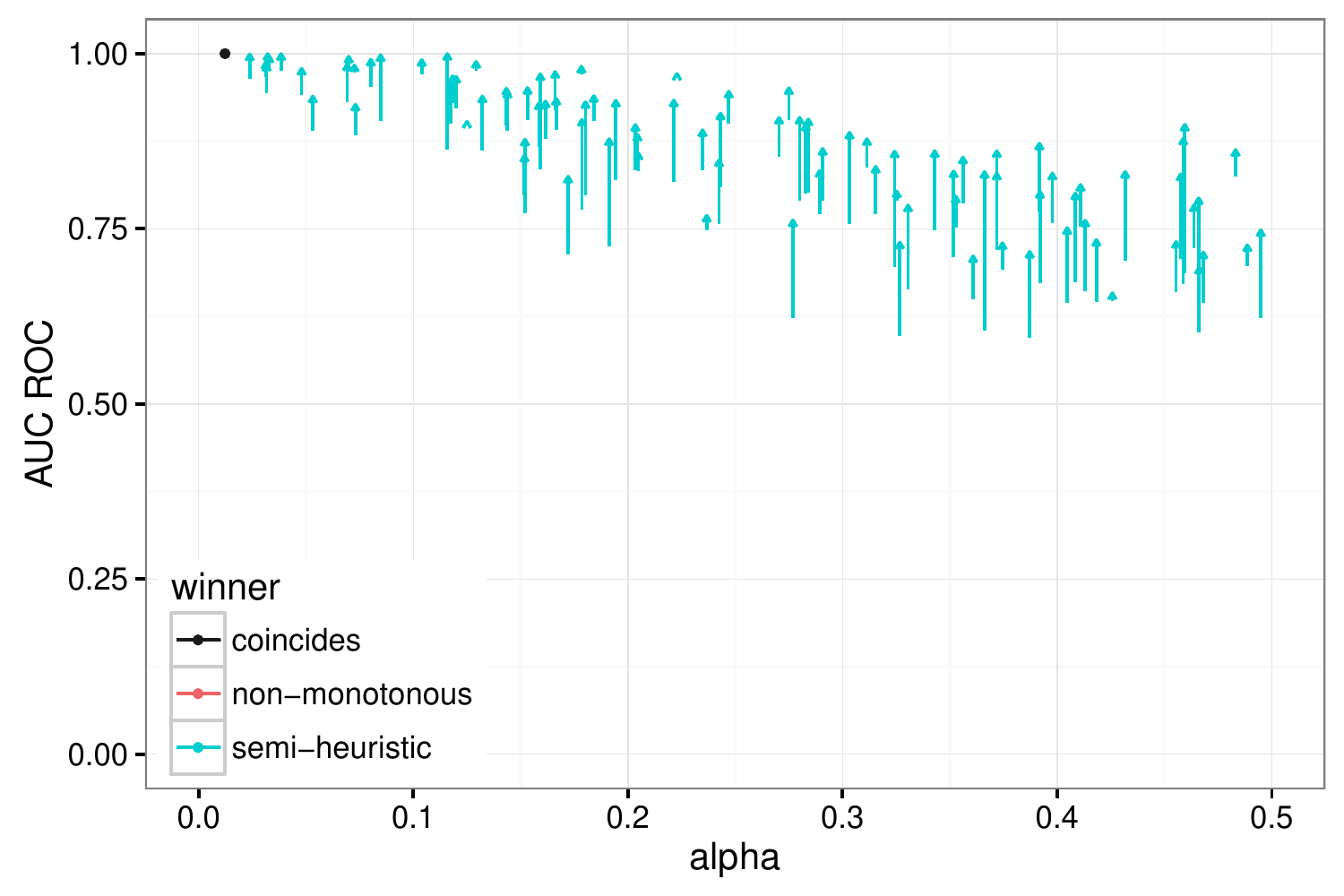} &
        \includegraphics[width=6cm]{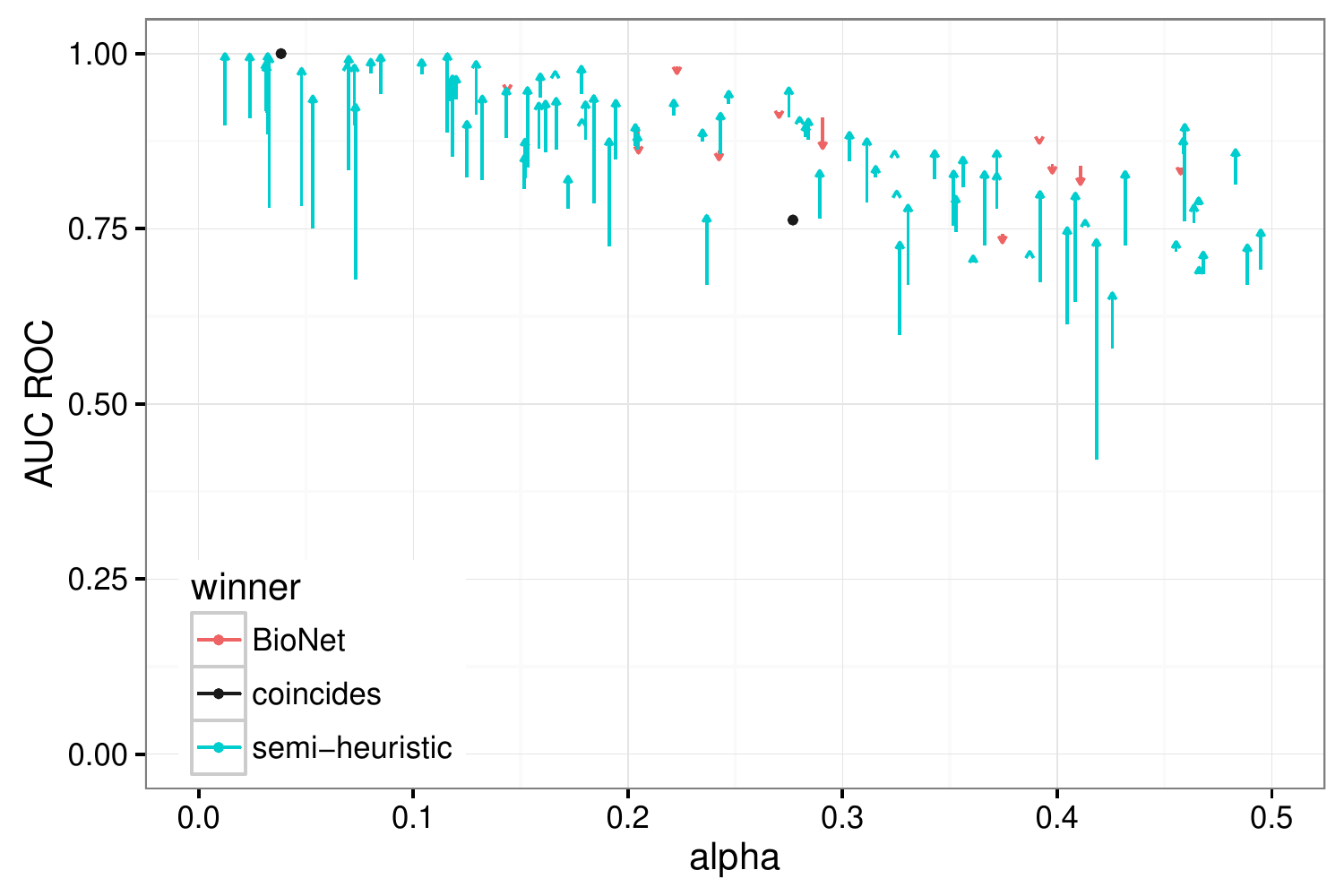}  
    \end{tabular}
    \caption{Module AUC values for graphs of size 100. Three methods
        are present: semi-heuristic, BioNet-like and non-monotonous. }%
    \label{fig:med}%
\end{figure}

The results of the experiment are shown on Fig.~\ref{fig:med}. 
Almost for all cases semi-heuristic ranking have worked better than
both BioNet-like and non-monotonous baseline methods.

\subsection{Large real-world graph}

Finally, we analyzed performance of the proposed semi-heuristic method
on the large real-word graph. For this experiment we used 
a protein-protein interaction graph from the example of BioNet package~\cite{Beisser2010}.
This graph has 2089 vertices and 7788 edges. An active module in this network
was sample to be a size of 50--250.

\begin{figure}
    \centering
    \begin{tabular}{@{}cccc@{}}
        \includegraphics[width=6cm]{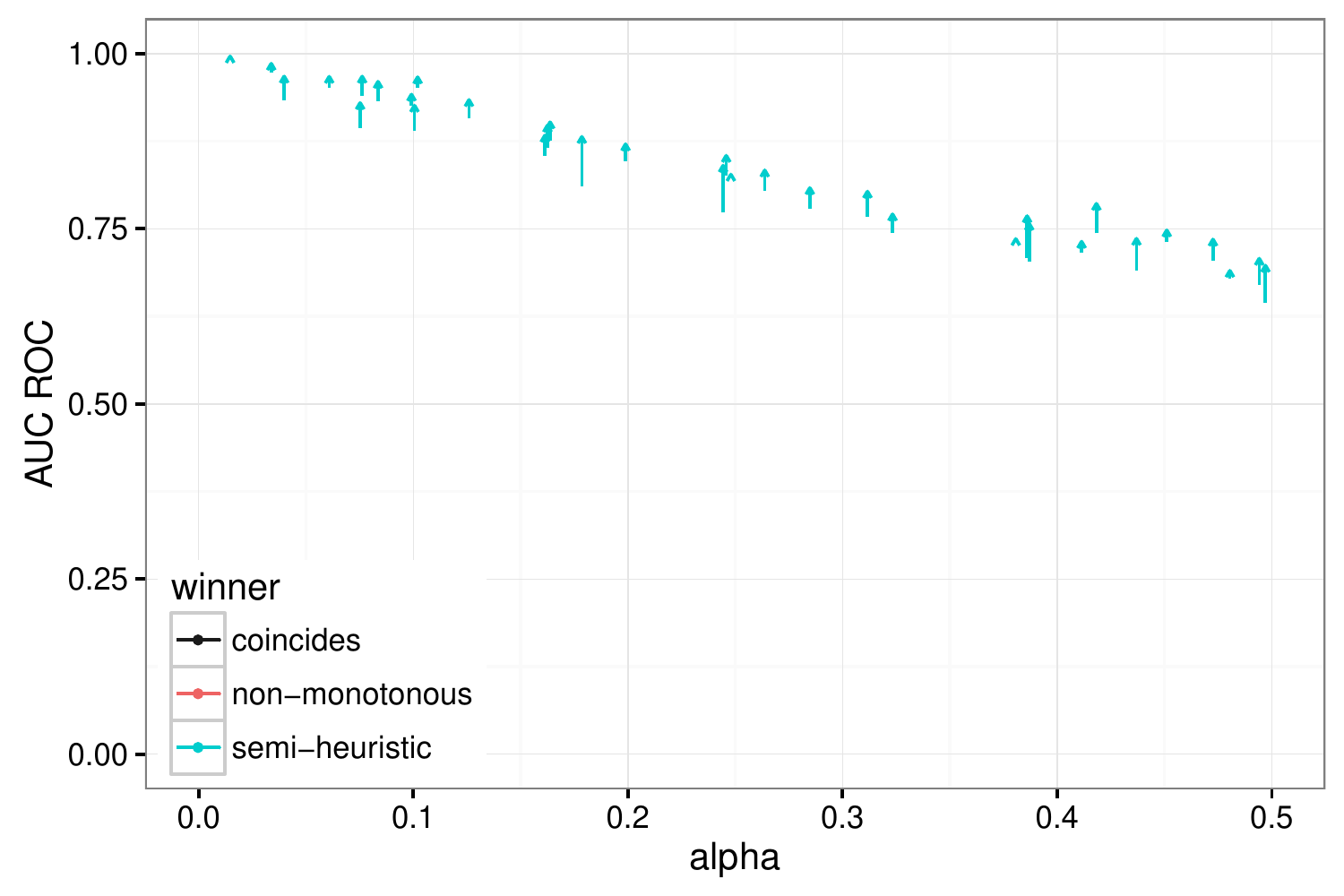} &
        \includegraphics[width=6cm]{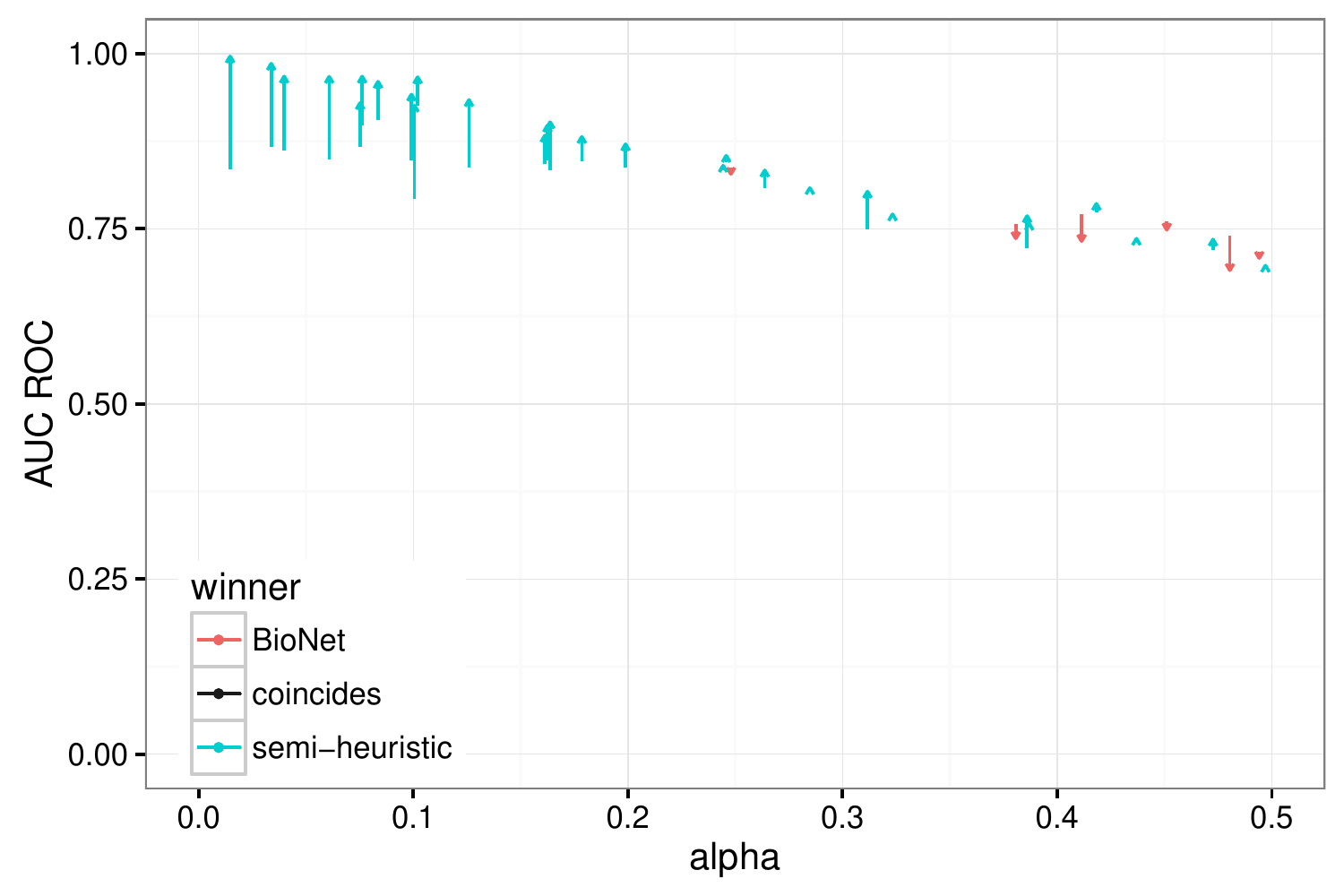}    
    \end{tabular}
    \caption{Module AUC values for a real protein-protein interaction graph.
        Three methods are present: semi-heuristic, BioNet-like and non-monotonous. }%
    \label{fig:lar}%
\end{figure}

The results of the experiment are shown on Fig.~\ref{fig:lar}. As for 
medium sizes semi-heuristic method works better than both baseline methods.
On the other hand, the running time of the method increased significantly 
to about six hours.

\subsection{Generating graphs for experiments}
\label{sec_details}

To mimic real network graphs generated for the experiments were scale-free.
For the generation we used an existing implementation of the Barabasi-Albert algorithm 
from an R-package igraph.

For subgraph sampling of the given size we used the following procedure.
Let $G=(V,E)$ be a connected graph,
$k$ be a required size of an active module and $M$ is the set of
vertices of the generated random active module. At the beginning $M$ is empty.
First we add into $M$ a random vertex from the graph.
Next we choose one of the adjacent vertex of $M$
that does not already belong to $M$ and add it. 
This step is repeated until $M$ is of size $k$.

\section{Conclusion}

The problem of active module recovery appears in many
areas of bioinformatics. Usually it is solved by an heuristic or exact
algorithm that provides a module for a selected significance threshold.
However, in practice multiple threshold values are tested and the
results of these tests are not easily combined to be interpreted.
In this paper we considered a ranking variant of this problem, where
vertices are ranked before a particular threshold is selected.
We also force a property of a module for a more stringent threshold to be
a subgraph of a module for a less stringent one.
We proposed two methods to solve this problem. The first method uses
dynamic programming to find a ranking that maximizes an expected value
of AUC score. We consider this method to be optimal, but it works only on small graphs.
The second method does not explicitly maximize the AUC score but
compares well to the optimal one and works better than the baseline 
methods in practice. However, it is also able to rank graphs with up to thousands vertices
in a reasonable time.

\bibliography{paper}{}
\bibliographystyle{splncs03}

\end{document}